\documentclass{emulateapj}

\usepackage{array}
\usepackage{times}

\slugcomment{The Astronomical Journal, 130:2424--2433, 2005 November}

\shorttitle{ADAPTIVE RFI CANCELLATION IN RADIO ASTRONOMY}
\shortauthors{MITCHELL, ROBERTSON, \& SAULT}

\begin{document}


\title{Alternative Adaptive Filter Structures for improved radio frequency interference Cancellation in Radio Astronomy}


\author{D. A. Mitchell,\altaffilmark{1,2}
        J. G. Robertson,\altaffilmark{1} and
        R. J. Sault\altaffilmark{3}}


\altaffiltext{1}{School of Physics, The University of Sydney, NSW 2006, Australia}
\altaffiltext{2}{Affiliated with the Australia Telescope National Facility, CSIRO.}
\altaffiltext{3}{CSIRO Australia Telescope National Facility, Epping, NSW 1710, Australia}


\begin{abstract}

In radio astronomy, reference signals from auxiliary antennae, receiving only
the radio frequency interference (RFI), can be modified to model the RFI
environment at the astronomy receivers. The RFI can then be cancelled from the
astronomy signal paths. However, astronomers typically only require signal
statistics. If the RFI statistics are changing slowly, the cancellation can be
applied to the signal correlations at a much lower rate than required for
standard adaptive filters. In this paper we describe five canceller setups;
pre- and post-correlation cancellers that use one or two reference signals in
different ways. The theoretical residual RFI and added noise levels are
examined, and demonstrated using microwave television RFI at the Australia
Telescope Compact Array. The RFI is attenuated to below the system noise, a
reduction of at least 20 dB. While dual-reference cancellers add more reference
noise than single-reference cancellers, this noise is zero-mean and only adds
to the system noise, decreasing the sensitivity. The residual RFI that remains
in the output of single-reference cancellers (but not dual-reference
cancellers) sets a non-zero noise floor that does not act like random system
noise and may limit the achievable sensitivity. Thus dual-reference cancellers
often result in superior cancellation. Dual-reference pre-correlation
cancellers require a double-canceller setup to be useful and to give equivalent
results to dual-reference post-correlation cancellers.

\end{abstract}


%%%%%%%%%%%%%%%%%%%%%%%%%%%%%%%%%%%%%%%%%%%%%%%%%%%%%%%%%%%%%%%%%%%%%%%%%%%%%%%%
%%%%%%%%%%%%%%%%%%%%%%%%%%%%%%%%%%%%%%%%%%%%%%%%%%%%%%%%%%%%%%%%%%%%%%%%%%%%%%%%
\section{INTRODUCTION}
\label{INTRODUCTION}

Radio astronomy is poised to move ahead with a suite of new instruments such as
the Square Kilometre Array (SKA). These instruments will improve sensitivity,
resolution, bandwidth, and many other instrument and observational parameters
by more than an order of magnitude. At present, many radio astronomy
observations are corrupted to at least some extent by radio frequency
interference (RFI). This interference comes from ground-based communication
transmitters, satellites, and the observatory equipment itself. With the
increase in sensitivity and frequency coverage of radio astronomy instruments,
and with telecommunication signals occupying more of the spectrum, it is
essential to develop ways of removing or suppressing this RFI.

In radio astronomy, real-time adaptive filters can be used to modify an
auxiliary voltage time series (the reference signal) so that it cancels RFI
from an astronomical voltage time series \citep{Barnbaum1998, Bower2001,
Mitchell2005}. For each voltage sample the filters are allowed to slightly vary
their internal coefficients in order to adapt to any changes taking place in
the RFI. If one is interested in the power spectrum of the astronomy signal,
and the filter coefficients stay fairly constant over the interval in which the
power spectrum is estimated, cancellation in the post-correlation domain can
give better results, since a second reference antenna can be used to give
complete suppression of the RFI (only zero-mean random receiver noise is added
to the complex correlations and it will average away, see \citealt{Briggs2000,
Mitchell2001}). However, many applications (particularly in the communications
field, but also some astronomy applications), require the recovery of the
actual symbol stream (i.e., a transmitted sequence of symbols such as bits or
words) from the noisy RF environment, which is not retained in
post-correlation. Following a suggestion of \citet{Briggs2000}, we have devised
a modified approach that can give improved RFI attenuation in the voltage
domain. In an attempt to minimise any RFI in an astronomical voltage series the
standard approach is to minimise the canceller's output power, which it will be
shown means that some residual RFI always remains. The modified approach that
we give here forces the RFI in the output power to zero. It results in residual
power that is always greater than that of the standard approach (output power
is no longer minimised), but which does not contain RFI. That is, superior RFI
cancellation is obtained, but at the expense of somewhat increased thermal
noise.

In the following sections the standard adaptive canceller is discussed, and the
new approach is introduced. This is followed by an overview of how cancellation
can be applied after correlations are formed. Residual RFI power and added
receiver noise are investigated and an example from the Australia Telescope
Compact Array is given.

%%%%%%%%%%%%%%%%%%%%%%%%%%%%%%%%%%%%%%%%%%%%%%%%%%%%%%%%%%%%%%%%%%%%%%%%%%%%%%%%
%%%%%%%%%%%%%%%%%%%%%%%%%%%%%%%%%%%%%%%%%%%%%%%%%%%%%%%%%%%%%%%%%%%%%%%%%%%%%%%%
\section{ADAPTIVE CANCELLERS}
\label{ADAPTIVE CANCELLERS}

%%%%%%%%%%%%%%%%%%%%%%%%%%%%%%%%%%%%%%%%%%%%%%%%%%%%%%%%%%%%%%%%%%%%%%%%%%%%%%%%
\subsection{The Model}
\label{THE MODEL}

In an attempt to remain general, we assume a system of one or more radio
antennae pointing towards a direction on the celestial sphere. Delays are
inserted into the signal paths so that a wavefront from the chosen direction
arrives at the output of each antenna simultaneously. The celestial location is
known as the phase tracking centre \citep{Thomson1986,Taylor1999}. Given that
we will attempt to deal with the interference in subsequent parts of the
system, separate reference antennae are incorporated into the network of
receivers to observe the RF environment, which typically enters the astronomy
signal through the side-lobes of the antennae. Signals from astronomy antennae
will be referred to as main signals, and those from reference antennae as
reference signals. At each antenna a waveform containing an additive mixture of
all the signals present in the environment is received and downconverted to an
IF voltage series. This is sampled and quantised into a number of digital bits
(which we assume is sufficient to keep the voltage statistics linear so that
quantisation effects such as intermodulation are negligible, and to ensure that
the receiver noise and astronomy fluctuations are measured even in the presence
of strong interference). Each main IF voltage series contains three components:
a noise voltage from the receiving system, $n(t)$; a noise voltage from the
sky, $s(t)$; and interference, $i(t)$. The sky voltage contains the information
about the astronomical sources, which are the signals of interest. If the
interference cannot be removed completely, it is desirable to reduce it to less
than the final RMS noise level (with a negligible -- or at least predictable --
effect on the astronomy, \citealt{Barnbaum1998}).

Since an interfering signal is usually incident from a direction other than the
phase tracking centre, its wavefront will not be synchronous at the output of
the different antennae. The geometric delay, $\tau_{k}$, of antenna $k$ -- due
to the physical separation of the receivers -- represents the difference in
arrival time of an interfering wavefront at antenna $k$ and an arbitrary
reference point (after accounting for the delay needed to track the selected
field on the celestial sphere).

As a signal passes through a receiving and processing system it encounters
various convolutions and deconvolutions, so working in the frequency domain can
offer a more intuitive basis for discussion. In the frequency domain the system
can be represented by complex multiplications and divisions. A
frequency-dependent coupling term, $G_{\nu}(t)$, is used to describe the
combined complex-valued gain of each receiver system and antenna to the
interference, including any filtering (time is included to account for the slow
variations imposed as the RFI passes through antenna side-lobes).  Using upper
case characters to denote frequency domain quantities, and keeping in mind that
this spectral representation comes from Fourier transforming each consecutive
1000 or so samples of the voltage series, the signal in a quasi-monochromatic
channel at frequency $\nu$ is

\begin{equation}
V_{\nu,m}=N_{\nu,m}+G_{\nu,m}(t)I_\nu e^{j\phi_{\nu,m}(t)}+S_\nu,\\
\end{equation}

\noindent
for main antennae and

\begin{equation}
V_{\nu,r}=N_{\nu,r}+G_{\nu,r}(t)I_\nu e^{j\phi_{\nu,r}(t)},\\
\end{equation}

\noindent for reference antennae. Note that the phase term due to the geometric
delay of the interfering signal, $\phi_{\nu,k}(t)=2\pi\nu\tau_{k}(t)$, has been
kept separate from the $G$-terms (time is also included here to account for
changes as either the phase tracking centre or RFI transmitter direction
change). To remain general the $G$-terms are kept as complex quantities, to
allow for any effects on the phase that are not due to the geometric delay. It
is assumed that there is only one interfering signal in a frequency channel
(see \citealt{Bower2001} for a discussion of multiple interferers), and that
there is negligible reference antenna gain in the direction of $S$, i.e.,
reference antennae do not measure signal from astronomical sources. These are
important assumptions, but often quite reasonable (the latter assumption is
strengthened because the weak astronomy signal enters the reference antennae
through side-lobes). However, statements made later in relation to the lack of
effect of adaptive cancellers on the astronomy signal rely on the validity of
the second assumption. If any astronomy signals leak into the reference series
the achievable sensitivity, dynamic range, astronomy purity, etc., will all be
affected. For example, signal leakage could lead to the power of a strong
self-calibration source changing as the canceller weights change.

%%%%%%%%%%%%%%%%%%%%%%%%%%%%%%%%%%%%%%%%%%%%%%%%%%%%%%%%%%%%%%%%%%%%%%%%%%%%%%%%
\subsection{MK1: Single Reference Adaptive Cancellers}
\label{SINGLE REFERENCE SIGNAL ADAPTIVE CANCELLERS}

Adaptive cancellers are usually applied to broadband IF voltage samples in the
time domain \citep{Widrow1985,Barnbaum1998}. (``Broadband'' here simply refers
to the whole passband, rather than the quasi-monochromatic frequency channels.)
While the adaptive canceller examples given throughout section \ref{AN ATCA
DATA EXAMPLE} have been processed in the time domain, the following analysis is
carried out in the frequency domain (see \citealp{Widrow1985,Barnbaum1998}; and
\citealp{Bower2001} for descriptions of time domain implementation).

The aim of adaptive cancellers in interference mitigation is to find the set of
filter weights, {\textbf{\textit{W}}}, which scale and phase shift each
reference antenna frequency channel so that they best approximate the RFI in a
main astronomy spectrum (in the time domain delays are inserted into the signal
paths so that positive and negative delays can be considered. So strictly
speaking the cancellers are not truly real-time, the output is lagging in time
by the length of the inserted delay.) So ${\textbf{\textit{W}}}$ is a vector
with a complex element for each frequency channel. Figure \ref{mk1 adaptive
cancellers} shows schematically how such a canceller can be implemented,
hereinafter referred to as a mark one (MK1a) canceller (the ``a'' is added
since we will be modifying the filter later).

%\clearpage

\begin{figure}[t]
  \centering
  %\plotone{f1.eps}
  \includegraphics[width=8.5cm]{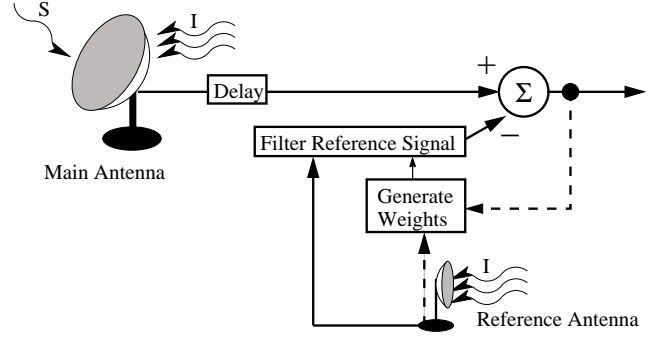}
  \caption{\small Schematic of a MK1a adaptive canceller. Dashed lines indicate
		  signal paths used to calculate the weights.}
  \label{mk1 adaptive cancellers}
\end{figure}

%\clearpage

Throughout this paper it is assumed that the filter weights are varying slowly
enough that they are approximately constant on the time scales used to
calculate them; typically less than a second. It is also assumed that the
various signals are independent (i.e., $S_{m}$, $I_{m}$, and $N_{m}$ are
uncorrelated, and the receiver noise is independent for the different antennae,
i.e., $N_{m}$ and $N_{r}$ are uncorrelated. The noise terms, $N$, will not be
independent for receivers used in low frequency instruments such as LOFAR,
where they will be dominated by partially correlated sky noise. Here we will
only consider frequencies for which $N$ is dominated by uncorrelated noise,
internal to the receivers.) When these assumptions hold, the power in a single
frequency channel at the output of the canceller is

\begin{equation}\begin{array}{rcl}
P & = & \left<(V_{m}-WV_{r})^{ }(V_{m}-WV_{r})^{*}\right>\\
&&\\
 & = & \left<|V_{m}|^2\right>-\left<V_{m}^{ }W^{*}V_{r}^{*}\right>-\\
 &   & \left<V_{m}^{*}W^{ }V_{r}^{ }\right>+\left<|WV_{r}|^2\right>,
\label{adaptive filter power}
\end{array}\end{equation}

\noindent where * denotes a complex conjugate and $\left<...\right>$ the
expectation value. The explicit frequency dependency of the terms has been
removed (it is assumed that the channels are independent so that (\ref{adaptive
filter power}) can be applied to each frequency channel separately). If we set
$\sigma_I^2=\left<|I|^2\right>$, $\sigma_S^2=\left<|S|^2\right>$,
$\sigma_{N_k}^2=\left<|N_{k}|^2\right>$, and assume that the complex gains,
delays, and weights are constant over the time average, so they can be taken
outside the average, then we can write

\begin{equation}\begin{array}{rcl}
P & = & \sigma_{N_m}^2+\sigma_S^2+
        |G_{m}|^2\sigma_I^2-\\
  &   &  W^{*}G_{m}^{ }G_{r}^{*}e^{ j\phi_{mr}}\sigma_I^2-
         W^{ }G_{m}^{*}G_{r}^{ }e^{-j\phi_{mr}}\sigma_I^2+\\
  &   & |W|^2|G_{r}|^2\sigma_I^2+
        |W|^2\sigma_{N_r}^2,
\label{mk1 weight dependence}
\end{array}\end{equation}

\noindent where $\phi_{mr}$ represents the phase difference,
$\phi_{m}-\phi_{r}$.

Equation (\ref{mk1 weight dependence}) highlights a critical point. Minimising
the output power reduces the \emph{combined} power of the residual RFI and the
inserted reference receiver noise. If there is reference receiver noise the RFI
will never be completely cancelled \citep{Widrow1985,Barnbaum1998}. This is
because both the interference and reference receiver noise are being weighted.
The receiver noise of the main antennae and the astronomy signals, however, are
not affected by the choice of weights and will pass through the canceller
freely (under assumptions of signal independence and zero reference gain
towards astronomy sources).

To perform the minimisation of $P$, one can differentiate it with respect to
$W$ and find the weights that set the derivative equal to zero. The surface of
$P$ is a multidimensional (positive) quadratic surface that has a single
minimum, so the weights that set $dP/dW=0$ must give the unique global minimum
\citep{Widrow1985,Barnbaum1998}. Alternatively, Wiener theory tells us that the
optimal weights (known as the Wiener-Hopf solution), $W_{\it{mk1a}}$, which
minimise the output power are also the weights that set the cross-correlation
between the canceller output and the reference signal to zero
(\citealt{Widrow1985}, as indicated in figure \ref{mk1 adaptive cancellers}):

\begin{eqnarray}
& \,
& \left<(V_{m}^{}-W^{}_{\it{mk1a}}V_{r}^{ })V_{r}^{*}\right> = 0,
  \label{mk1 pre corr weights a}\\
%\nonumber\\
%\addtocounter{equation}{-1}\addtocounter{subequation}{1}
\Rightarrow &\,&
\displaystyle
W_{\it{mk1a}} = \frac{\left<V_{m}^{ }V_{r}^{*}\right>}
                     {\left<V_{r}^{ }V_{r}^{*}\right>}.
  \label{mk1 pre corr weights b}
\end{eqnarray}

So the weighting process takes the cross-correlation of the reference and main
signals, which determines the correlated power and relative delay of the
interference, and scales that by the auto-correlation of the reference signal.
One can calculate these weights directly by calculating the correlations from
short integrations, or they can be found adaptively by iteratively seeking and
then tracking the weights that satisfy (\ref{mk1 pre corr weights a}).

Ideally the output of the canceller would consist of the astronomy signal and
receiver noise. However, as mentioned above, since there is always some
receiver noise in the reference signal, setting the correlation in (\ref{mk1
pre corr weights a}) to zero can never remove all of the RFI. The RFI is played
off against the reference receiver noise. If $\it{INR}_r$ is the
interference-to-noise power ratio of the reference signal,
$G_{r}^{2}\sigma_I^2/\sigma_{N_r}^2$, the mean amount of residual output power
(power in addition to the main receiver noise and the astronomy,
$R_{\it{mk1a}}=P_{\it{mk1a}}-\sigma_{N_m}^2-\sigma_S^2$), is given by

\begin{equation}
R_{\it{mk1a}} = \frac{G_{m}^{2}\sigma_I^2}{1+\it{INR}_{r}}.
\label{mk1 residual power}
\end{equation}

It is clear from (\ref{mk1 residual power}) that as $\it{INR}_r$ approaches
infinity (no reference receiver noise) the residual power goes to zero. If
there is no RFI, $G_{m}^{2}$ will be zero and there will also be zero residual
power (the filter turns off). When $\it{INR}_r$ is finite and non-zero the
reference receiver noise term in the denominator of (\ref{mk1 pre corr weights
b}) biases the weights and some residual power will remain. What might not be
so clear from an inspection of (\ref{mk1 residual power}) is the statement made
earlier that this residual power is a combination of reference receiver thermal
noise added during cancelling and residual RFI that was not excised. Another
way to see this bias is to consider figure \ref{mk1 adaptive cancellers}.
Thermal noise from the reference receiver is present in both inputs to the
weight generation process (in the $V_{r}$ term from the reference antenna and
the $-W_{\it{mk1a}}V_{r}$ term from the filter output, c.f. equation \ref{mk1
pre corr weights a}). This will lead to a second non-zero-mean correlation
product (the first being the RFI). Minimising the output power must be a
trade-off between minimising the contributed reference receiver noise and the
RFI, and as a result there will always be residual RFI. The total residual
power given in (\ref{mk1 residual power}) can be divided into the inserted
reference receiver noise residual, $R_{\it{\it{mk1a}}}^{\it{(rx)}}$, and the
RFI residual, $R_{\it{\it{mk1a}}}^{\it{(rfi)}}$, such that

\begin{equation}
R_{\it{\it{mk1a}}} = R_{\it{\it{mk1a}}}^{\it{(rfi)}} +
                     R_{\it{\it{mk1a}}}^{\it{(rx)}},
\end{equation}

\noindent which are shown by \citealt{Mitchell2004} to be

\begin{equation}
R_{\it{mk1a}}^{\it{(rfi)}}
 = \frac{G_{m}^{2}\sigma_I^2}
        {\it{INR}_{r}^{2}\left(1+\it{INR}_{r}^{-1}\right)^{2}},
\label{mk1 residuals a}
\end{equation}

\begin{equation}
R_{\it{mk1a}}^{\it{(rx)}}
 = \frac{G_{m}^{2}\sigma_I^2}
        {\it{INR}_{r}^{ }\left(1+\it{INR}_{r}^{-1}\right)^{2}}.
\label{mk1 residuals b}
\end{equation}

Thus $R_{\it{mk1a}}^{\it{(rfi)}} = R_{\it{mk1a}}^{\it{(rx)}}\,/\,\it{INR}_{r}$.
To shed some light on the meaning of the relations in (\ref{mk1 residuals a})
and (\ref{mk1 residuals b}), we again interpret the process of output power
minimisation as determining the weights that set the cross-correlation of the
canceller output and the reference signal to zero. As $\it{INR}_r$ increases,
the RFI becomes the dominant signal in the cross-correlation, and the filter
must concentrate on reducing the RFI power. As a result the proportion of the
RFI that remains after cancelling decreases faster than the injected noise
power. When $\it{INR}_r<1$ noise starts to dominate and the filter will
concentrate on reducing the contributed noise power. When $\it{INR}_r$ goes to
zero, the correlation is completely reference receiver noise, and the canceller
turns itself off. When RFI dominates, most of the residual power is reference
receiver noise, and when reference receiver noise dominates, most of the
residual power is RFI. Also, since the reference signal is being scaled in an
attempt to match its own RFI to the RFI in the main signal, the larger $G_{r}$
is relative to $G_{m}$ (for example using a reference antenna that is pointing
directly at the interfering source), the smaller the scaling factor (weighting
amplitude) and thus the amount of injected receiver noise. When $\it{INR}_r \gg
1$ and $G_{r} \gg G_{m}$ both of the residual terms drop off and extremely good
results are achieved \citep{Barnbaum1998,Bower2001}.

%%%%%%%%%%%%%%%%%%%%%%%%%%%%%%%%%%%%%%%%%%%%%%%%%%%%%%%%%%%%%%%%%%%%%%%%%%%%%%%%
\subsection{MK2: Dual Reference Adaptive Cancellers}
\label{DUAL REFERENCE SIGNAL ADAPTIVE CANCELLERS}

To remove the biasing effect caused by the reference receiver noise, a second
reference receiver can be used, as was suggested in \cite{Briggs2000}. The RFI
in the main spectrum is still estimated using a weighted version of the
spectral channels from the first reference, but now the cross-correlation of
the second reference signal with the canceller output is set to zero in order
to find the weights. This MK2a canceller is shown in figure \ref{mk2 adaptive
cancellers}.

%\clearpage

\begin{figure}[ht]
  \centering
  %\plotone{f2.eps}
  \includegraphics[width=8.5cm]{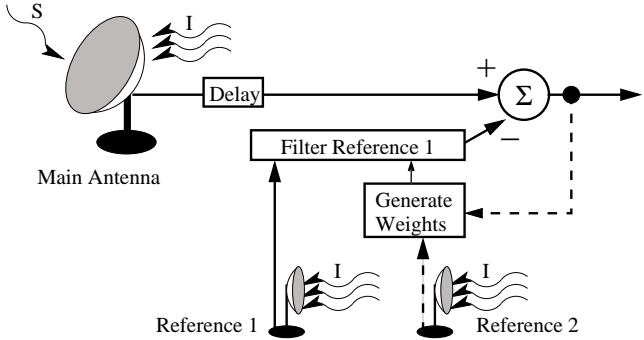}
  \caption{\small Schematic of a MK2a adaptive canceller, which uses two
		  reference signals. Dashed lines indicate signal
		  paths used to calculate the weights.}
  \label{mk2 adaptive cancellers}
\end{figure}

%\clearpage

It was mentioned in the previous section that the reference receiver noise is a
component of both signals in (\ref{mk1 pre corr weights a}). Zeroing this
correlation results in minimum output power, but a non-zero RFI residual.
Correlating the canceller output against a second reference with uncorrelated
receiver noise removes the bias. Since the RFI is the only correlated signal,
zeroing the cross-correlation results in zeroing the RFI. However, there is no
built-in mechanism to guard against the amount of reference receiver noise
contributed to the output. The amount added during cancelling must always be
greater than the noise added by the MK1a canceller, since power is no longer
minimised.

The two references will be denoted $r_1$ and $r_2$, so that

\begin{equation}\begin{array}{rcl}
V_{r_1} & = & N_{r_1}+G_{r_1}Ie^{j\phi_{r_1}}\mbox{, and}\\
\\
V_{r_2} & = & N_{r_2}+G_{r_2}Ie^{j\phi_{r_2}},\\
\end{array}\end{equation}

\noindent and we can again determine the weights that set the output-reference
cross-correlation to zero:

\begin{equation}\begin{array}{rl}
&
\left<(V_{m}^{}-W_{\it{mk2a}}^{}V_{r_1}^{})V_{r_2}^{*}\right>=0\\
&\\
\Rightarrow &
\displaystyle
W_{\it{mk2a}} = \frac{\left<V_{m}^{}V_{r_2}^{*}\right>}
                     {\left<V_{r_1}^{}V_{r_2}^{*}\right>}.
\label{mk2 pre corr weights}
\end{array}\end{equation}

As with the MK1a canceller the weights need to scale and phase shift reference
signal $V_{r_1}$ so that its RFI component matches that of signal $V_{m}$.
Here, however, the independent reference signal $r_2$ is used to give a true
view of the scaling factor needed to match RFI levels. $W_{\it{mk2a}}$ is only
constrained by the RFI, since the receiver noise terms in $\left<V_{r_1}^{
}V_{r_2}^{*}\right>$ are uncorrelated, and the RFI signal is entirely replaced
with a weighted version of thermal noise from reference signal $V_{r_1}$
(assuming that the various signals are uncorrelated and that the receivers are
ideal and remain linear). While the noise in the weights will also increase the
output power of the canceller, it is much weaker noise since the weights are
averaged over many samples, and is not considered further here. When $W_{mk2a}$
from (\ref{mk2 pre corr weights}) is substituted into the equation for output
power, $P_{mk2a}=\left<|V_{m}-W_{mk2a}V_{r_1}|^2\right>$, the mean residual
power is

\begin{equation}
R_{\it{mk2a}}
 = R_{\it{mk2a}}^{\it{(rx)}}
 = \frac{G_{m}^{2}\sigma_I^2}{\it{INR}_{r_1}},
\label{mk2 residual power}
\end{equation}

\noindent as shown in \citet{Mitchell2004}. Infinite attenuation of the RFI
component has been achieved, but potentially a significant amount of system
noise has been added during cancelling. Compare (\ref{mk1 residual power}) and
(\ref{mk2 residual power}): As in (\ref{mk1 residual power}), when
$\it{INR}_{r_1}\gg 1$, the attenuation of the RFI signal in $V_{m}$ is very
large. In this case however, any frequency channels with $\it{INR}_{r_1}< 1$
will end up with more unwanted power than they started with. The MK2a canceller
does not turn itself off. The reference signal is boosted until its RFI matches
the RFI in the main signal, regardless of the amount of receiver noise being
added. Note that $\it{INR}_{r_2}$ does not affect the output power, provided
there is enough RFI power in $V_{r_2}$ to keep the weights stable (see section
\ref{INSTABILITIES IN THE DUAL REFERENCE ALGORITHMS} for a discussion of weight
stability).

If one is interested in the (interference-free) power spectrum of $V_{m}$, the
single reference antenna MK1a canceller output will contain less residual
power. There are, however, advantages to the MK2a canceller. If one is
concerned with retrieving a structured signal from a voltage series, random
noise in the signal may not pose too much of a problem, but a structured RFI
residual may detract from signal recovery. More relevant in radio astronomy is
the case where one is looking for a structure in the power spectrum. The RFI
remaining after MK1a cancelling will have features in the power spectrum, but
if the (amplified and filtered) reference RFI, $G_{r_1}^{2}\sigma_I^2$, is
proportional to $G_{m}^{2}\sigma_I^2$, equation (\ref{mk2 residual power}) says
that the noise contributed by MK2a cancelling will have the same spectrum as
the input reference receiver noise (apart from a constant scaling factor). It
is also conceivable to remove the unwanted reference receiver noise from the
auto-correlation of either canceller's output. This is discussed in the next
section and can result in a MK2 canceller superior to the MK1 canceller for
some applications.

%%%%%%%%%%%%%%%%%%%%%%%%%%%%%%%%%%%%%%%%%%%%%%%%%%%%%%%%%%%%%%%%%%%%%%%%%%%%%%%%
\subsection{Suppressing Added Reference Noise}
\label{AVERAGING OUT REFERENCE RECEIVER NOISE ADDED DURING CANCELLING}

We now describe a double canceller setup that can be used to suppress the added
reference receiver noise in the output astronomy power spectrum. If the main
signal is duplicated before cancelling so there is an identical copy, the copy
can be passed through a second canceller that uses different reference signals
so that the noise added will be uncorrelated with the noise added to the first.
When the two filtered copies are cross-correlated, the main antenna receiver
noise and astronomy will correlate as if the original signal has simply been
auto-correlated, while the added reference noise power will average away with
the radiometric factor ($\sqrt{\Delta\nu\tau_{\it{int}}}$).

\begin{figure}[ht]
  \centering
  %\plotone{f3.eps}
  \includegraphics[width=8.5cm]{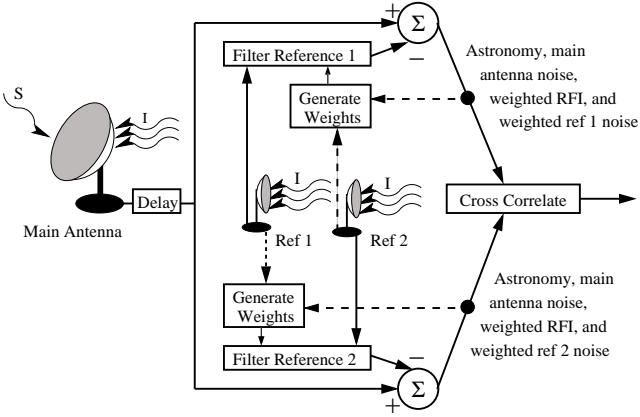}
  \caption{\small MK2b canceller system where the auto-correlation is
		  replaced by the cross-correlation of two independently
		  processed copies of the main signal.}
  \label{indepRX MK2}
\end{figure}

An important point to note about the MK2a canceller is that all of the noise
added during cancelling is from the first reference receiver (see equation
\ref{mk2 residual power}). The second reference is only used in setting the
complex weights. So the second canceller can be made by interchanging the
references. This setup will be called the MK2b canceller and is shown in figure
\ref{indepRX MK2}. As discussed in \cite{Mitchell2004}, if the INR of the
references are the same, the mean residual output power from the MK2b canceller
is

\begin{equation}
R_{\it{mk2b}} = \frac{R_{\it{mk2a}}}{\sqrt{\Delta\nu\tau_{\it{\it{int}}}}},
\label{mk2 indepRX}
\end{equation}

\noindent which averages towards zero as the integration length is increased.

%\clearpage

%\clearpage

\begin{figure}[ht]
  \centering
  %\plotone{f4.eps}
  \includegraphics[width=8.5cm]{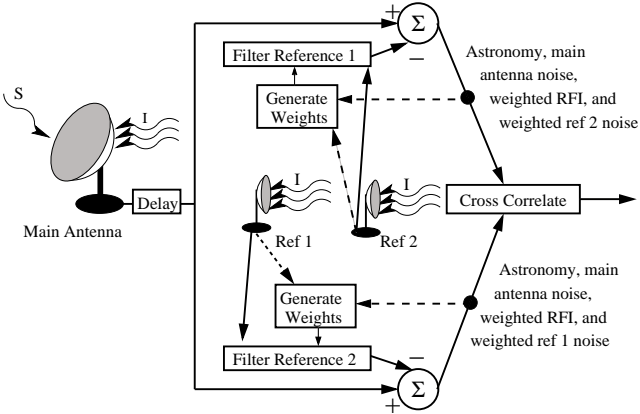}
  \caption{\small MK1b canceller system where the auto-correlation is
		  replaced by the cross-correlation of two independently
		  processed copies of the main signal.}
  \label{indepRX MK1}
\end{figure}

Similarly, if two references are used to create two independent MK1a cancellers
for the two main signal copies, this gives the MK1b canceller shown in figure
\ref{indepRX MK1}, and the mean residual output power becomes

\begin{equation}
R_{\it{mk1b}}
  = R_{\it{mk1a}}^{\it{(rfi)}} +
    \frac{R_{\it{mk1a}}^{\it{(rx)}}}{\sqrt{\Delta\nu\tau_{\it{int}}}}.
\label{mk1 indepRX}
\end{equation}

%\clearpage

%\clearpage

Equations (\ref{mk2 indepRX}) and (\ref{mk1 indepRX}) show that while
$R_{\it{mk2b}}$ should keep integrating towards zero, $R_{\it{mk1b}}$ has a
definite limit due to the RFI signal that remains after cancelling. However,
one must be aware that in situations where the reference INR is very small the
MK2 canceller does not turn itself off, and there are practical implementation
issues that need to be addressed (essentially, one may need to force the
canceller to turn off). This is discussed in section \ref{INSTABILITIES IN THE
DUAL REFERENCE ALGORITHMS}.

%%%%%%%%%%%%%%%%%%%%%%%%%%%%%%%%%%%%%%%%%%%%%%%%%%%%%%%%%%%%%%%%%%%%%%%%%%%%%%%%
\section{THEORETICAL RESULTS}
\label{THEORETICAL RESULTS}

In this section the residual power equations given throughout section
\ref{ADAPTIVE CANCELLERS} are demonstrated. Note that while the plots for the
dual-reference cancellers show the residual power as the reference
interference-to-noise power ratio approaches arbitrarily close to zero, the
algorithms in practice become unstable and need to be turned off. This point
will be reiterated where appropriate in the discussion below. For the theory we
have set $\sigma_S^2 = \sigma_{N_m}^2 = 0$, so that all of the output power
displayed in this section is a combination of residual RFI and any reference
receiver noise added during cancelling.

\begin{figure}[ht]
  \centering
  %\plotone{f5.eps}
  \includegraphics[width=8.5cm]{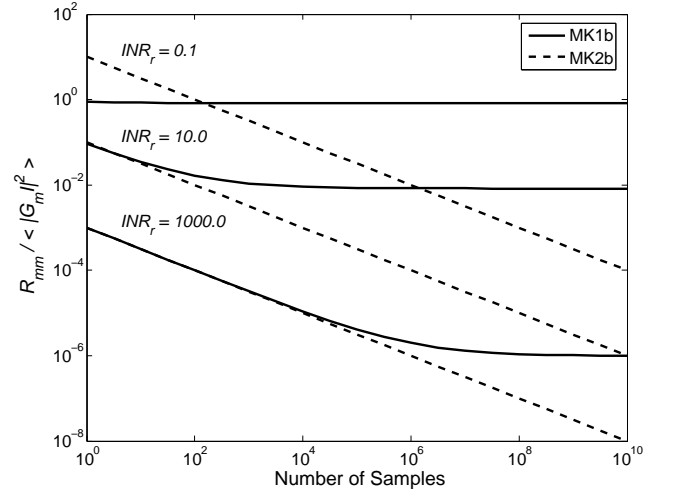}
  \caption{\small Theoretical residual power (normalised by the main signal's
		  input RFI power) for various $\it{INR}_r$ values as a
		  function of the number of samples used to average the noise
		  down in the canceller output ($\Delta\nu\tau_{\it{int}}$).
		  The three sets of solid and dashed lines are for $\it{INR}_r$
		  values of 0.1 (top), 10 (middle), and 1000 (bottom). Solid
		  lines indicate MK1b cancelling, dashed lines MK2b cancelling.
		  The lines flatten when the residual RFI power level is
		  reached.}

  \label{residual power v Btau 1}
\end{figure}

Figure \ref{residual power v Btau 1} displays the proportion of residual power
in the output signal after adaptive cancelling with MK1b and MK2b cancellers.
The plot shows that the added reference receiver noise averages away as the
integration length is increased. If single canceller systems were being
considered, then since any noise added during cancelling is sent to an
auto-correlator, the output power would remain constant (i.e., remain at the
levels shown on the left hand side of figure \ref{residual power v Btau 1}). It
is clear that the MK1b canceller hits a limit when it reaches the residual RFI,
but that the MK2b does not.

As $\it{INR}_r$ decreases the normalised output power of the MK1b canceller
levels off at 1, so there is no cancelling taking place. On the other hand the
MK2 canceller continues to insert more and more reference receiver noise in an
attempt to match the reference RFI to the main signal RFI. Even though the MK2b
canceller always has the larger total residual power, it is entirely zero-mean
noise and averages out with the radiometric factor. Again the reader should 
note that for low $\it{INR}_r$ values the MK2 canceller can become unstable and
requires an additional mechanism to turn off.

\begin{figure*}[ht]
  \centering
  \includegraphics[width=15.6cm]{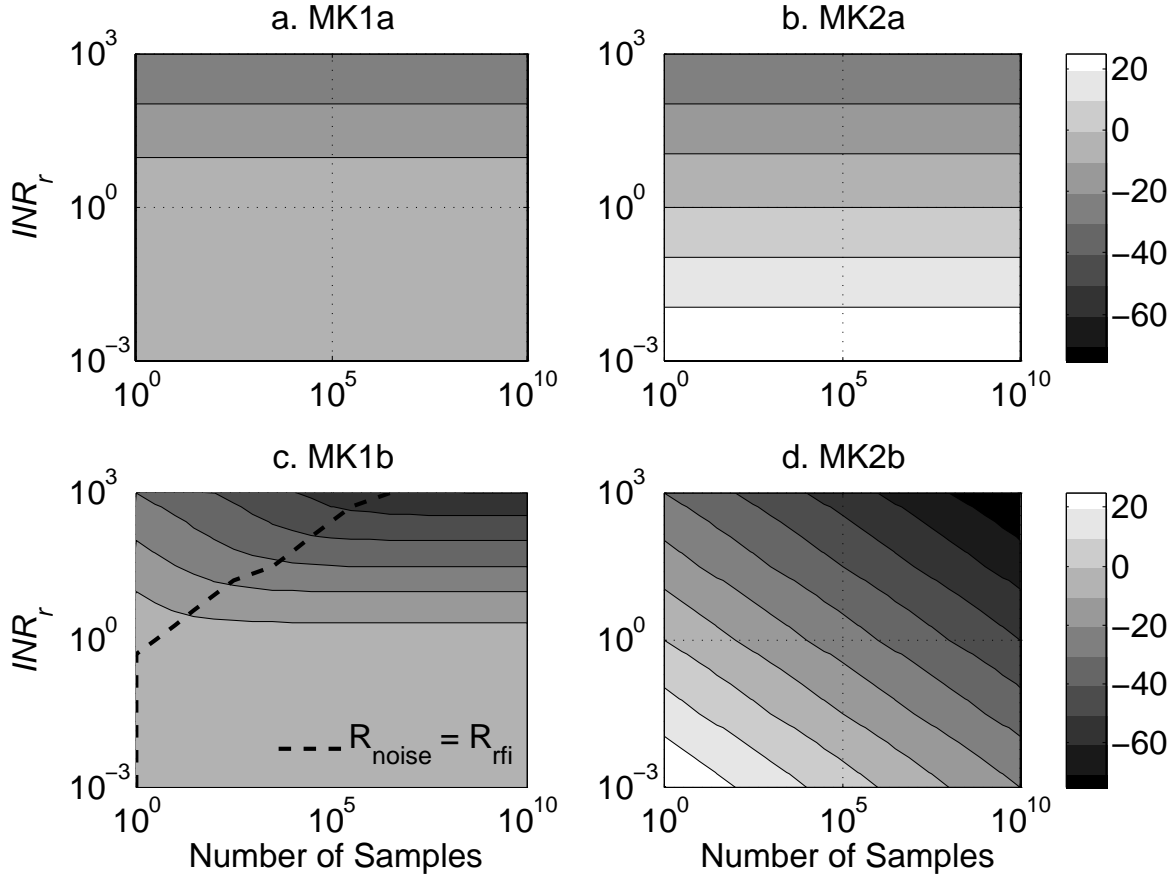}
  \caption{\small Residual power contours as a function of $\it{INR}_r$ and
		  $\Delta\nu\tau_{\it{int}}$. Greyscale indicates the
		  normalised output power, $R_{mm}/(|G_{m}|^{2}\sigma_I^2)$,
		  expressed in dB. Figures a and b show MK1a and MK2a
		  cancelling respectively. Figures c and d show MK1b and MK2b
		  canceller output respectively.}
  \label{residual power v INR}
\end{figure*}

Figure \ref{residual power v INR} shows contours of constant (normalised)
output power as a function of $\it{INR}_r$ and the number of samples,
$\Delta\nu\tau_{\it{int}}$. Figures \ref{residual power v INR}a and
\ref{residual power v INR}b represent MK1a and MK2a cancelling respectively,
and figures \ref{residual power v INR}c and \ref{residual power v INR}d
represent MK1b and MK2b cancelling respectively. The amount of residual power
in dB is indicated by the grey scale and runs from -80 to 20 dB. It is clear
from figure \ref{residual power v INR}c that a constant RFI residual remains
for all $\it{INR}_r$ values after MK1b cancelling. The dashed line indicates
the approximate line where the added reference receiver noise power has
averaged down to expose the non-zero residual RFI power level.

It is clear from figures \ref{residual power v INR}a and \ref{residual power v
INR}c that an output power plateau is reached as $\it{INR}_r$ becomes small for
the MK1 cancellers. The residual power of the plateau is 0 dB, and indicates
that the canceller has turned off. In contrast, the MK2a canceller
(\ref{residual power v INR}b) does not turn off and results in more output
power than input RFI power for low $\it{INR}_r$ levels. However, since the
residual is entirely noise it averages down in a MK2b canceller, where two
independent filters are used (\ref{residual power v INR}d). This is highlighted
further in the next section.

%%%%%%%%%%%%%%%%%%%%%%%%%%%%%%%%%%%%%%%%%%%%%%%%%%%%%%%%%%%%%%%%%%%%%%%%%%%%%%%%
\section{MICROWAVE RFI AT THE ATCA}
\label{AN ATCA DATA EXAMPLE}

We now demonstrate adaptive cancellation of real RFI impinging on the Australia
Telescope Compact Array (ATCA). The RFI is a point-to-point microwave (MW)
television link transmitted from a TV tower on a nearby mountain at 1503 MHz.
The reference antennae were two orthogonal linearly polarised receivers on a
small reference horn pointed in the direction of the MW transmitter, as
described in \citet{Bell2001}.\footnote{Dataset srtca02.} A linearly polarised
receiver on a regular ATCA antenna, pointing at the sky and receiving the
microwave link interference through the antenna side-lobes, was used to collect
the main signal. The RFI is polarised, and as long as all three receivers are
at least partially polarised in the same sense as the RFI the cancellers will
work correctly (assuming that the polarisation cross-talk between the reference
receivers is negligible so the receiver noise is independent). The received
voltages were filtered in a 4MHz band centred at 1503 MHz, downconverted, and
sampled with 4-bit precision.

\begin{figure}[ht]
  \centering
  %\plotone{f7.eps}
  \includegraphics[width=8.5cm]{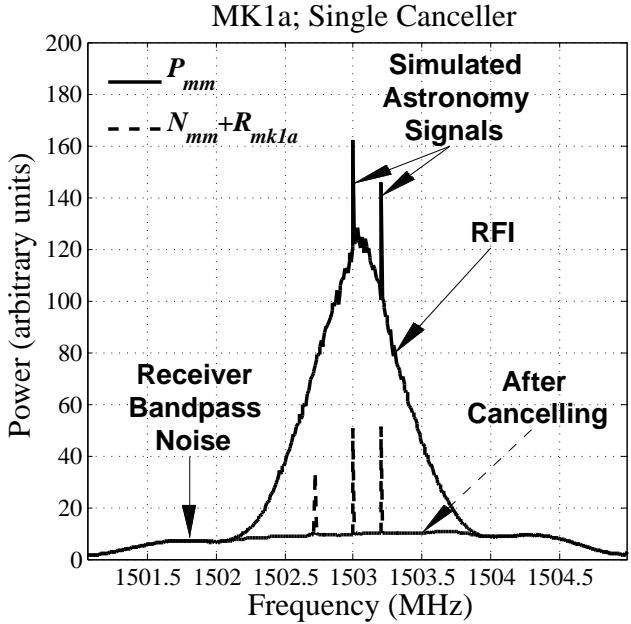}
  \caption{Power spectra of main signal $V_{m}$ before and after MK1 cancelling
	   with $INR_r\gg1$. Two simulated cosmic signals were added at 1503.0
	   and 1503.2 MHz.}
  \label{high INR mk1a spectrum}
\end{figure}

Each of the cancellation techniques discussed has been applied to the MW data
using MATLAB (see the \citealt{MATLAB1998}). All of the spectra shown were
generated using 1024-point FFTs. Figure \ref{high INR mk1a spectrum} shows the
unfiltered and MK1a filtered power spectra of the main astronomy voltage series
when large reference INRs were available. Two synthetic astronomy signals have
been added to the astronomy voltage series at 1503.0 MHz and 1503.2 MHz. What
we want is to remove the RFI peak (the 1-2 MHz wide peak centred at 1503 MHz)
from the main signal spectrum, leaving the broad, main antenna bandpass, and
the astronomy. This is indeed what is seen, and similar results are achieved
for all of the techniques discussed in this paper. Since all of the techniques
behave excellently when the reference INR is large, and any contributed
reference receiver noise or residual RFI is much smaller than the main signals
receiver noise level, it is difficult to compare them.

\begin{figure*}[ht]
  \centering
  \includegraphics[width=8.0cm]{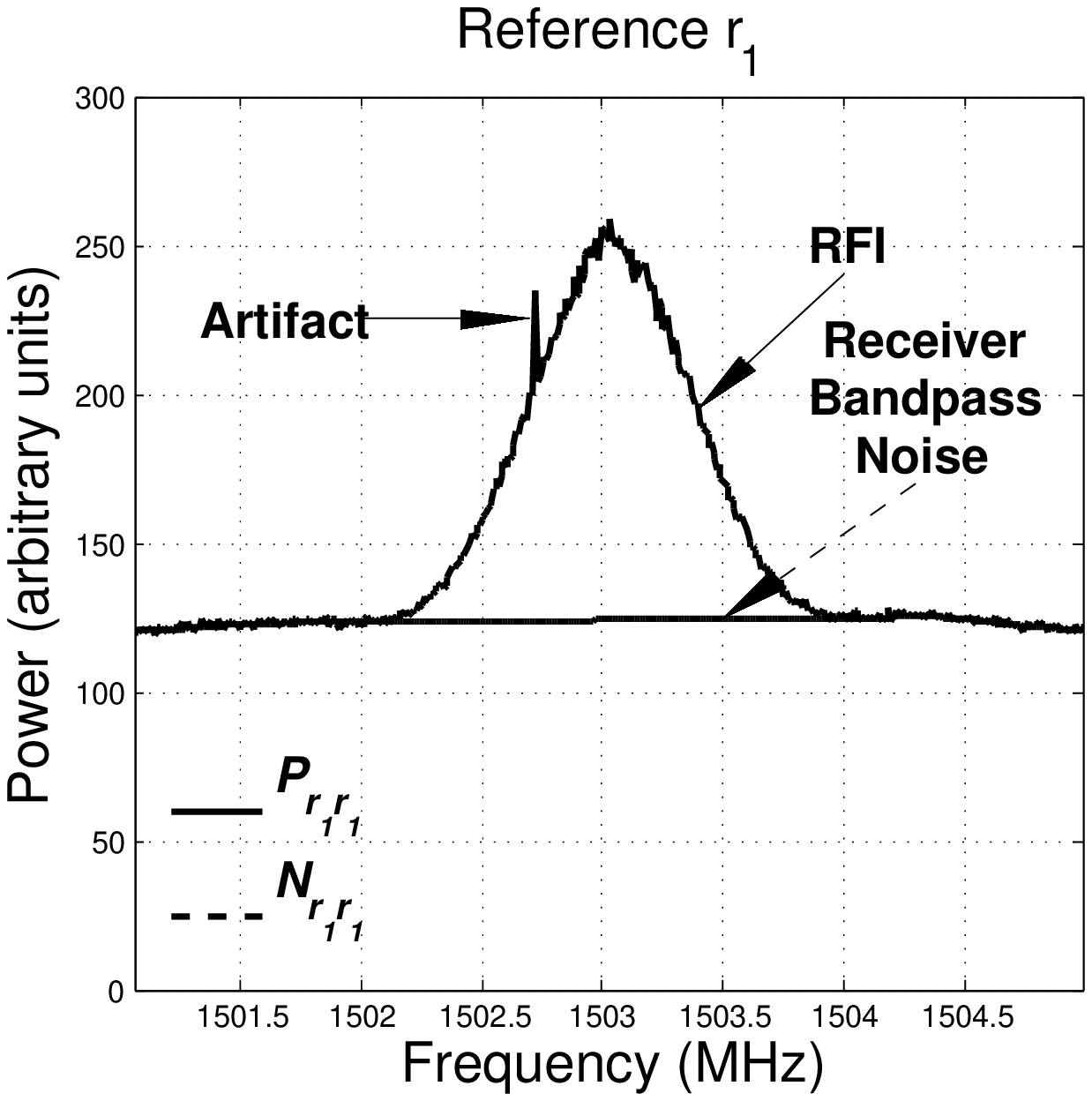}
  \includegraphics[width=8.0cm]{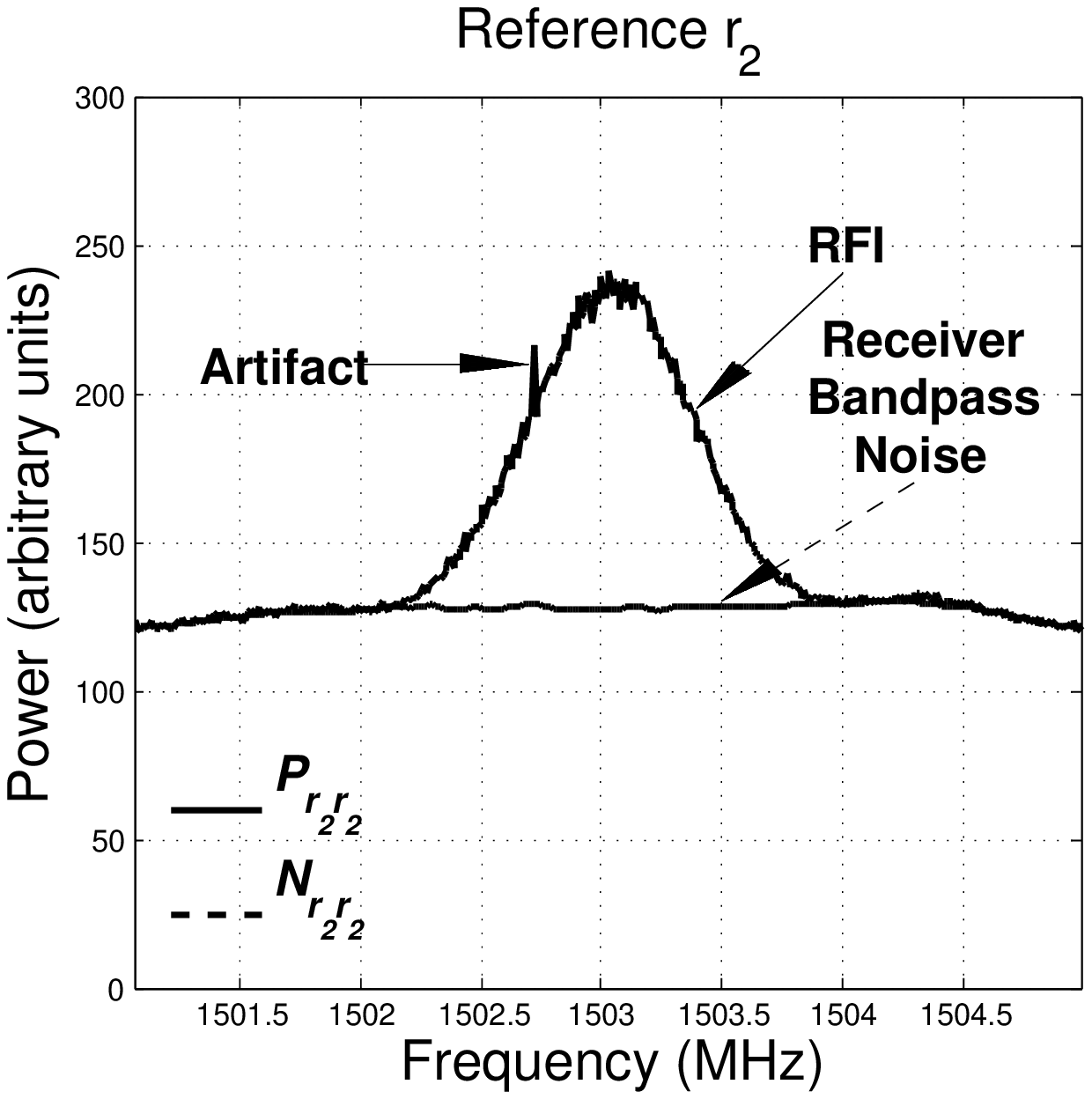}
  \vspace{5pt}
  \hbox{\hspace{1cm} (a) \hspace{7cm} (b)}
  \caption{\small Power spectra and estimated receiver noise of the reference
		  signals. Most of the bandpass noise was artificially added to
		  lower the INR. The peak at 1502.7 MHz is an artifact of the
		  receiving system that leaked into the reference voltages.
		  (a) Reference signal $V_{r_1}$.
		  (b) Reference signal $V_{r_2}$.}
  \label{reference spectra}
\end{figure*}

Figure \ref{reference spectra} displays the power spectra of two reference
signals that have had Gaussian random noise injected into their sampled voltage
series, to set $\it{INR}_r\approx1$ at the centre of the MW band and zero at
the edges. The reason for adding the fake receiver noise was to lower the INR
and accentuate any residual RFI, as detailed in (\ref{mk1 residuals a}),
(\ref{mk1 residuals b}), and (\ref{mk2 residual power}). Plots of the power
spectra of the main signal (the one containing the astronomy), before and after
the different cancellation techniques, are displayed in figures \ref{mk1
spectra} through \ref{post-corr spectra}.

\begin{figure*}[ht]
  \centering
  \includegraphics[width=8.0cm]{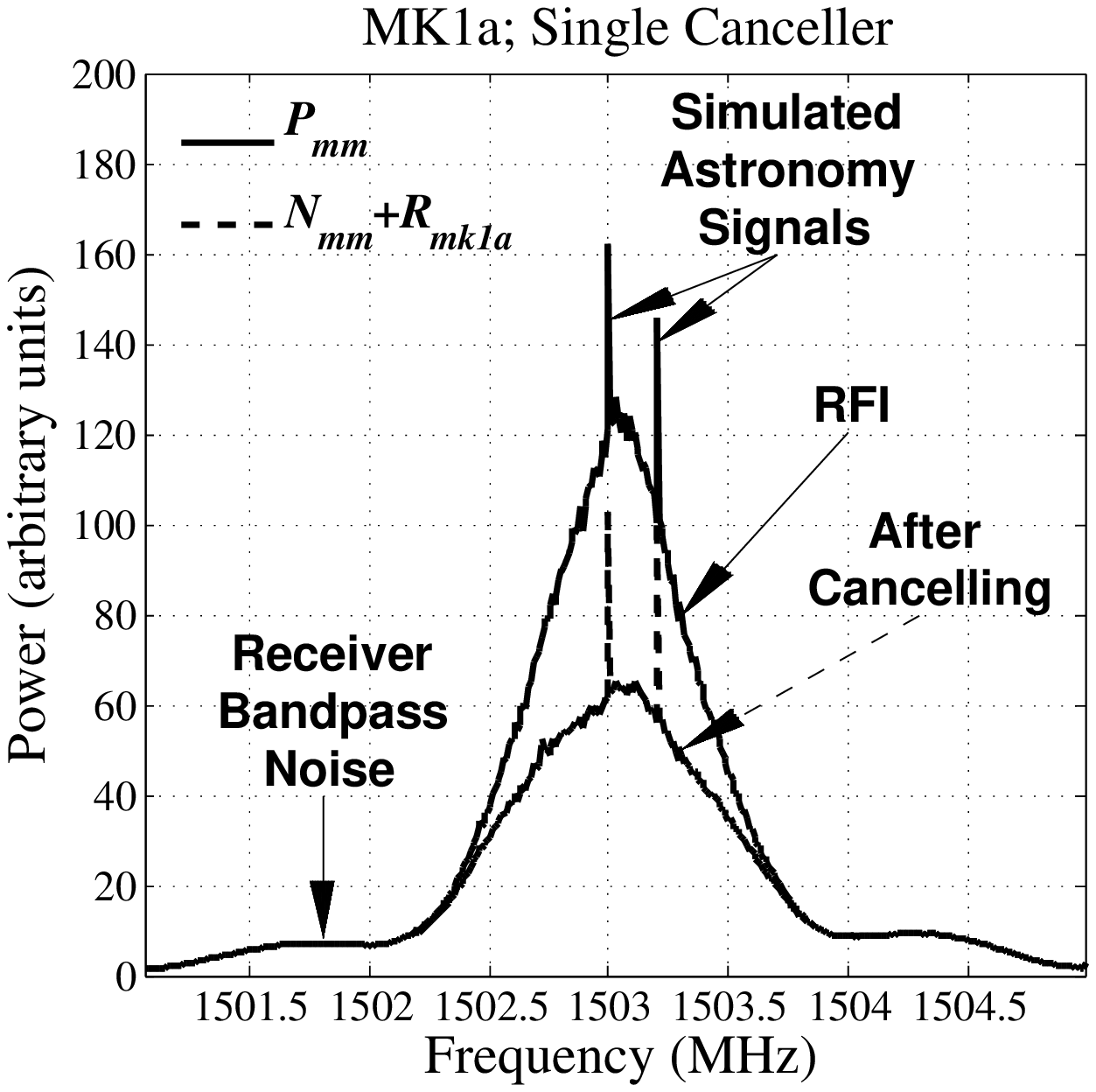}
  \includegraphics[width=8.0cm]{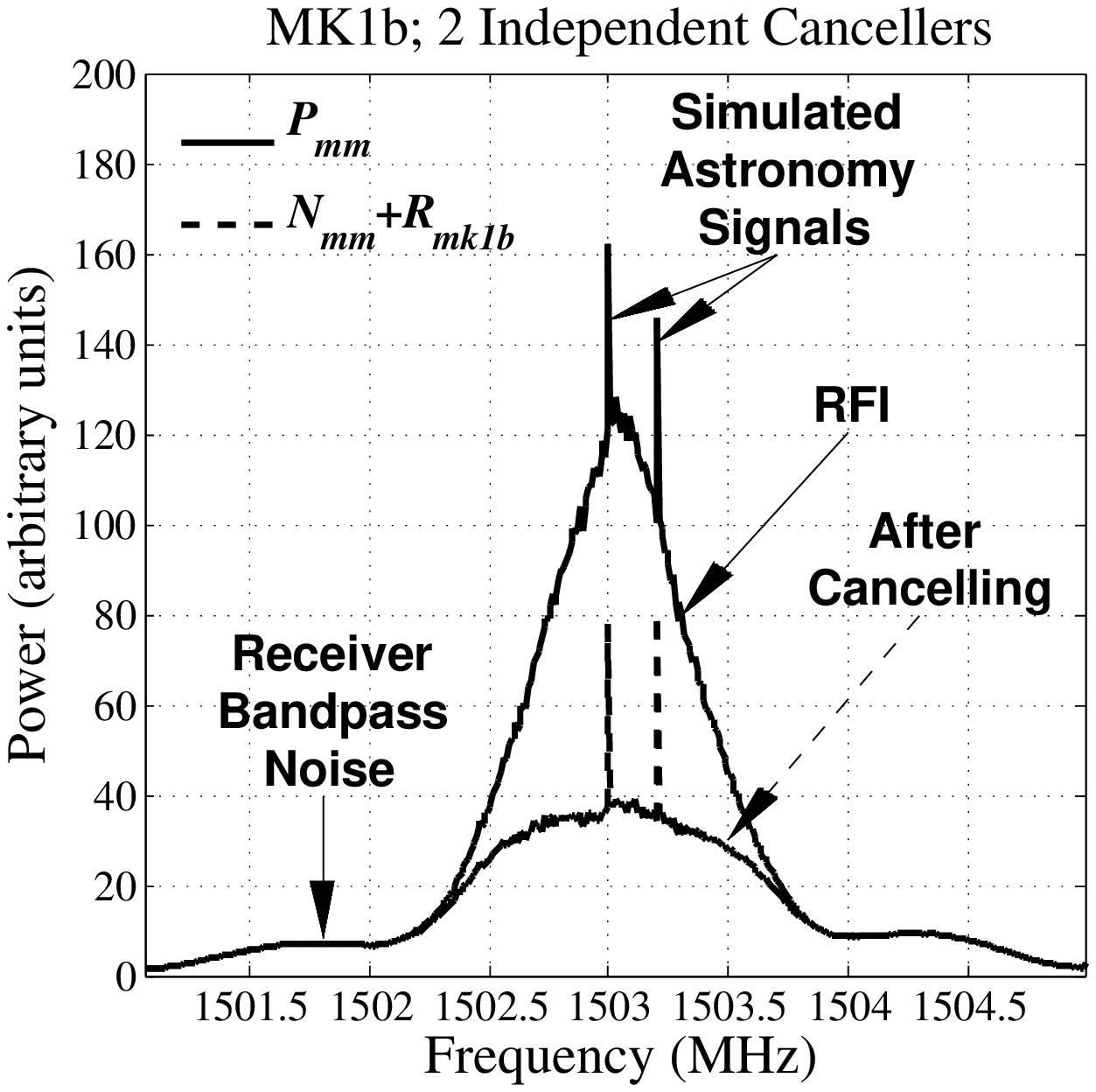}
  \vspace{5pt}
  \hbox{\hspace{1cm} (a) \hspace{7cm} (b)}
  \caption{\small Power spectra of main signal $V_{m}$ before and after MK1
		  cancelling. Two simulated cosmic signals were added at 1503.0
		  and 1503.2 MHz.
		  (a) MK1a canceller. (b) MK1b canceller.}
  \label{mk1 spectra}
\end{figure*}

Figure \ref{mk1 spectra} shows unfiltered and MK1 filtered power spectra of the
main astronomy voltage series. The two synthetic astronomy peaks have not been
affected to any measurable level. The amount of residual power after MK1a
cancelling is as expected from (\ref{mk1 residual power}). About twice as much
power has been cancelled at the centre of the RFI peak by the MK1b canceller
(the added receiver noise in this case has a zero-mean cross-correlation),
which is expected since $\it{INR}_r\approx1$. Away from the centre of the
peak, $\it{INR}_r$ becomes smaller and the proportion of residual RFI increases
(which is why the residual RFI peak is flatter than the initial peak). No
matter how long the integration is run this residual power will not decrease
any further. That is, the canceller is operating in the horizontal region shown
in figure \ref{residual power v Btau 1}.

\begin{figure*}[ht]
  \centering
  \includegraphics[width=8.0cm]{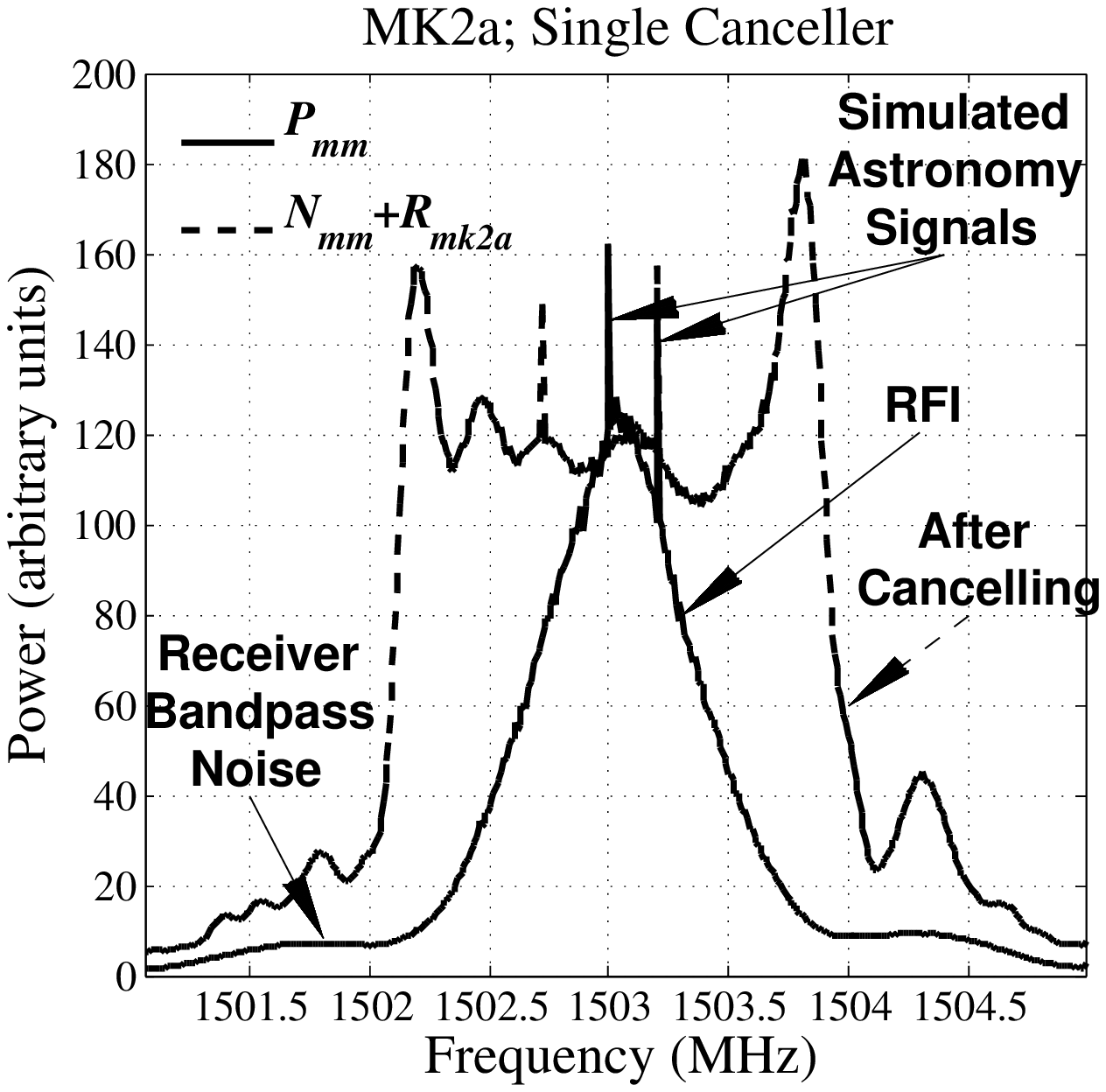}
  \includegraphics[width=8.0cm]{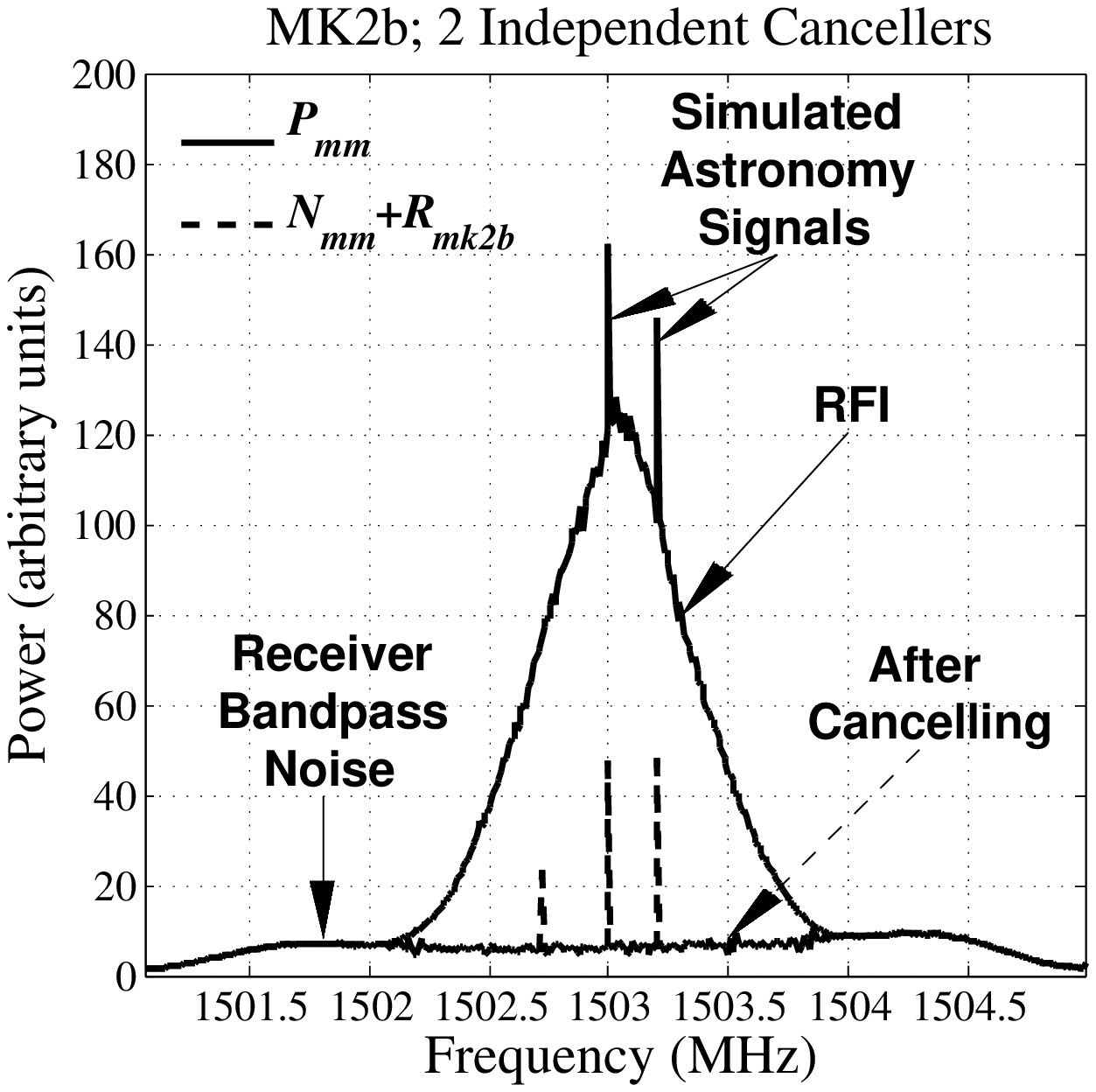}
  \vspace{5pt}
  \hbox{\hspace{1cm} (a) \hspace{7cm} (b)}
  \caption{\small Power spectra of main signal $V_{m}$ before and after MK2
		  cancelling. Two simulated cosmic signals were added at 1503.0
		  and 1503.2 MHz.
		  (a) MK2a canceller. (b) MK2b canceller.}
  \label{mk2 spectra}
\end{figure*}

\begin{figure}[ht]
  \centering
  %\plotone{f14.eps}
  \includegraphics[width=8.5cm]{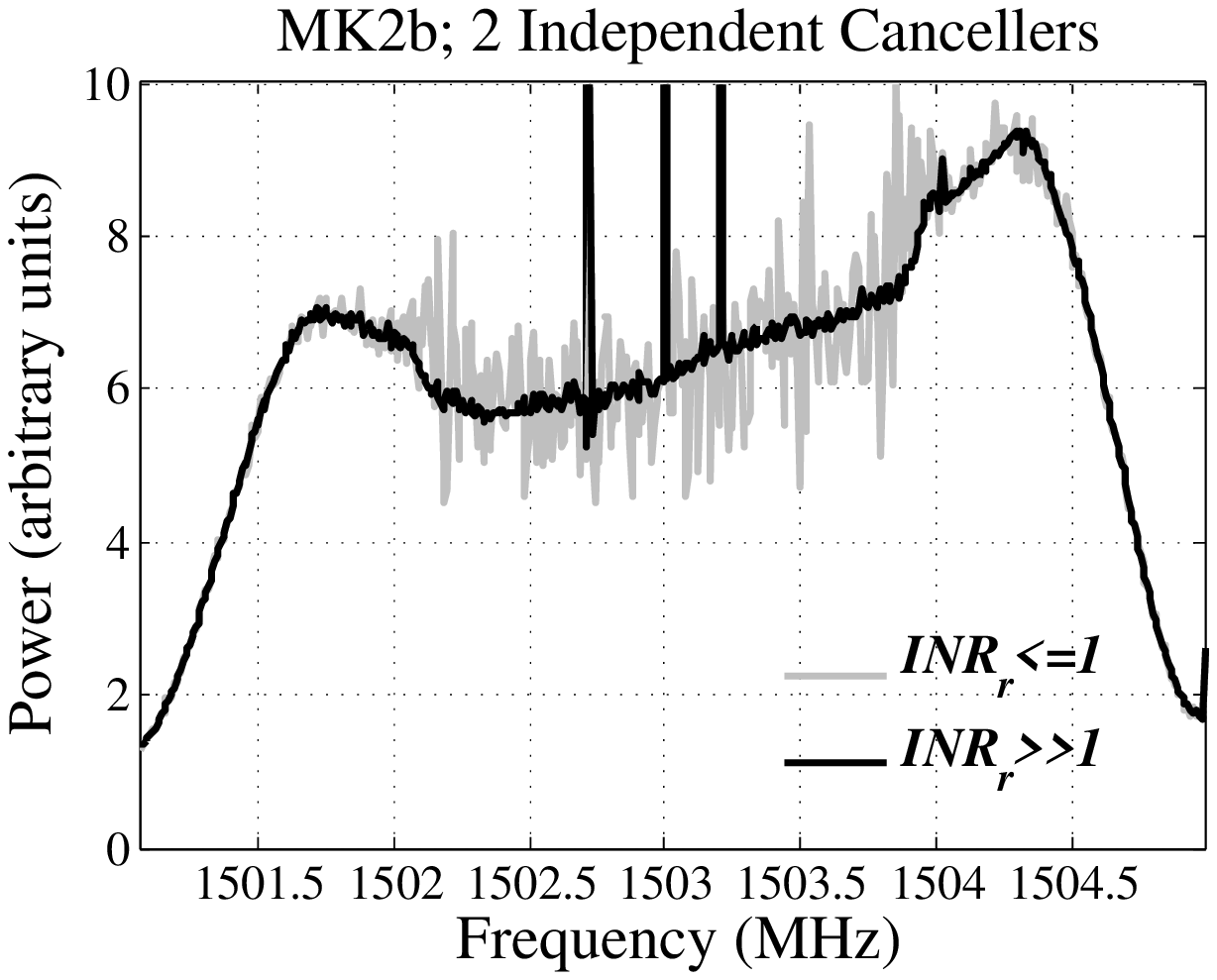}
  \caption{MK2b canceller spectrum from figure \ref{mk2 spectra}b, with a
	   $\it{INR}_r\gg1$ MK2b spectrum overlaid. Very little noise is added
	   when $\it{INR}_r\gg1$, demonstrating the increase for the
	   $\it{INR}_r<1$ case.}
  \label{mk2 spectra noise}
\end{figure}

Before and after spectra for the MK2 cancellers are shown in figure \ref{mk2
spectra}. As for the MK1 cancellers, the two synthetic astronomy signals have
been left unaffected. It is clear from figure \ref{mk2 spectra}a that when
$\it{INR}_r<1$ the MK2a output residual power is greater than the input RFI
power (except when $\it{INR}_r$ goes to zero, where the canceller was turned
off, see section \ref{INSTABILITIES IN THE DUAL REFERENCE ALGORITHMS}). This is
clearly unsuitable and makes the situation worse. In contrast, the MK2b
canceller (figure \ref{mk2 spectra}b) has removed the RFI peak down to the
primary antenna's receiver noise level, indicating that the MK2a residual power
was completely added noise and not residual RFI. The removal of the RFI was
extremely successful, with only a slight increase in zero-mean noise at the
primary receiver noise floor, as seen in figure \ref{mk2 spectra noise}.

Each spectral channel in the examples in this section was averaged for 15872
samples. This means that for the MK1b and MK2b cancellers the contributed
reference receiver noise power will have averaged down by about a factor of
$\sqrt{15872}\approx126$, or about 20 dB. Since the MK2b canceller output has
essentially the same noise spectrum as the MK2a canceller output, but averaged
down, there has been about a 20 dB reduction in unwanted power after MK2b
cancelling at the centre of the RFI peak (see figure \ref{mk2 spectra}). This
zero-mean noise would have continued to average down had we continued to
integrate, and more attenuation would have been achieved. The MK1b canceller
has only achieved around 6 dB of attenuation, and this will not increase with a
longer integration. One should keep in mind, however, that we have inserted a
substantial amount of noise into the reference voltages, which has also greatly
lowered the achieved attenuation.

An important final point to note is that, with the exception of the edges of
the RFI peak (where there are features associated with turning the filter off),
figure \ref{mk2 spectra}a shows that the RFI is spread with essentially
constant power over the RFI contaminated channels. This is because
$\it{INR}_r$, which we recall is given by $G_{r}^{2}\sigma_I^2/\sigma_{N_r}^2$,
is approximately proportional to $G_{m}^{2}\sigma_I^2$, so that the spectral
shape of the RFI cancels out of the residual power given in (\ref{mk2 residual
power}).

%%%%%%%%%%%%%%%%%%%%%%%%%%%%%%%%%%%%%%%%%%%%%%%%%%%%%%%%%%%%%%%%%%%%%%%%%%%%%%%%
\section{POST-CORRELATION CANCELLERS}
\label{POST-CORRELATION CANCELLERS}

From an implementation point of view it is important to realise that in
astronomy we are not usually seeking fast time-variable information such as
modulation. In most astronomical applications the aim is to measure signal
statistics since they are related to quantities such as cosmic flux density and
visibilities as measured by arrays. These applications are generally either
finding the auto-correlation of signals from a single antenna (to measure the
power spectrum of the astronomy signal), or the cross-correlation of signals
from more than one antennae (to measure the spatial coherence of the astronomy
signal). See \citet{Thomson1986} and various chapters of \citet{Taylor1999} for
an overview. In this paper we have concentrated on the power spectra given by
auto-correlations, however all of the techniques discussed can be generalised
to work on cross-correlations.

The upshot of only requiring signal statistics is that if the canceller weights
are not changing appreciably over the 100 millisecond or so time interval that
the statistics are measured over, then the algorithms can be applied to the
statistics rather than each voltage series. This means that they are applied at
a rate of Hz to kHz rather that MHz to GHz. It also means that if the new
astronomical cross-correlators that are coming online for new and existing
facilities have a few extra inputs for reference antennae, then no new filters
need to be added to the signal paths, and the cancellation can be performed
after the observation as part of post-processing. For a comparison of voltage
cancellation and statistics cancellation for RFI signals that require filters
with weights that are changing appreciably during an integration, see
\citet{Mitchell2005}.

The standard statistics canceller used in radio astronomy is a post-correlation
version of the MK2b canceller discussed in section \ref{DUAL REFERENCE SIGNAL
ADAPTIVE CANCELLERS}. If we use $P_{jk}$ to denote the correlation between
signals from antennae $j$ and $k$, the quantity that is subtracted from the
auto-correlation of antenna $m$ is determined from amplitude and phase closure
relations \citep{Briggs2000}:

\begin{equation}
C_{\it{pc2}} = \frac{P^{}_{mr_1}P^{*}_{mr_2}} { P^{*}_{r_1r_2} }
       \approx |G_{m}|^{2}\sigma_I^2.
\label{CORRECTION SPECTRUM MK2}
\end{equation}

\begin{figure}[ht]
  \centering
  %\plotone{f15.eps}
  \includegraphics[width=8.5cm]{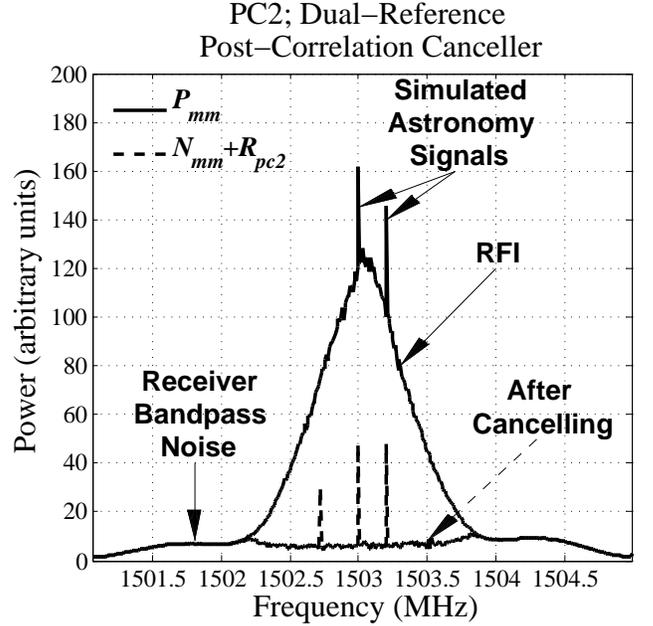}
  \caption{Power spectra of main signal $V_{m}$ before and after
	   post-correlation cancelling. Two simulated cosmic signals were added
	   at 1503.0 and 1503.2 MHz. There are slight artifacts at the edge of
	   the peak where $\it{INR}_{r}\rightarrow0$ and the canceller is
	   turning off, as described in section \ref{INSTABILITIES IN THE DUAL
	   REFERENCE ALGORITHMS}.}
  \label{post-corr spectra}
\end{figure}

Since the reference signal does not contain any information about the astronomy
signal, it will not be present in any of the quantities in (\ref{CORRECTION
SPECTRUM MK2}). The denominator removes the reference signal RFI's phase and
amplitude information from the numerator, leaving information about the RFI in
the signal from antenna $m$, and zero-mean noise, since the expectation
operators are not infinite in extent. When $C_{\it{pc2}}$ is subtracted from
main signal power $P_{mm}$, the mean residual power is (see
\citealt{Mitchell2004})

\begin{equation}
R_{\it{pc2}} = \frac{G_{m}^{2}\sigma_I^2}
                    {\sqrt{\it{INR}_{r_1}\it{INR}_{r_2}}
	             \sqrt{\Delta\nu\tau_{\it{int}}}},\\
\label{pc2 residual power}
\end{equation}

\noindent as in (\ref{mk2 indepRX}). $R_{\it{pc2}}$ is shown in figure
\ref{post-corr spectra}. As long as the gain and geometric delay stay
essentially constant over the time average, the ``pre'' and ``post''
correlation techniques are very similar; $R_{\it{pc2}}\approx R_{\it{mk2b}}$.

While in the last few sections an extreme case has be demonstrated (that of a
low reference INR), it does highlight major differences between the different
cancellers. The main differences being that infinite RFI removal is
theoretically possible for cancellers that use two reference signals, but not
for single reference cancellers. It is also clear that when the weights are
slowly varying, the post-correlation canceller is equivalent to the
dual-reference MK2b canceller.

Another way in which the MK2b and post-correlation canceller techniques can
differ is in their instabilities at low INR levels. Equations (\ref{mk2 pre
corr weights}) and (\ref{CORRECTION SPECTRUM MK2}) have denominators that
become zero-mean noise when there is no correlated RFI signal, which can lead
to numerical errors. The coefficients of the single-reference techniques go to
zero when the RFI becomes weak and the cancellers automatically turn off. These
issues are discussed next.

%%%%%%%%%%%%%%%%%%%%%%%%%%%%%%%%%%%%%%%%%%%%%%%%%%%%%%%%%%%%%%%%%%%%%%%%%%%%%%%%
\section{INSTABILITIES IN THE DUAL REFERENCE ALGORITHMS}
\label{INSTABILITIES IN THE DUAL REFERENCE ALGORITHMS}

Equations (\ref{mk2 pre corr weights}) and (\ref{CORRECTION SPECTRUM MK2}) show
that the dual-reference cancellers can have stability problems when
$\it{INR}_r\ll 1$. In frequency channels where the correlated interference is
zero or very small, (\ref{mk2 pre corr weights}) and (\ref{CORRECTION SPECTRUM
MK2}) are noise dominated and can result in a division by zero (or very close
to zero). This cannot occur in the MK1 cancellers since (\ref{mk1 pre corr
weights b}) goes to zero as $I\rightarrow0$ and they turn off. A modified
post-correlation canceller has been suggested by \citet{Briggs2000} in which an
extra term is added to the denominator of (\ref{CORRECTION SPECTRUM MK2}):

\begin{equation}
C_{\it{pc2}}^{\prime} = \frac{ P^{}_{mr_1}P^{*}_{mr_2}P^{}_{r_1r_2} }
          	             { \Psi^2 + P^{*}_{r_1r_2}P^{}_{r_1r_2} },
\label{MODIFIED SPECTRUM MK2}
\end{equation}

\noindent where $\Psi=\left(\left<N^{}_{r_1}N^{*}_{r_1}\right>
\left<N^{}_{r_2}N^{*}_{r_2}\right>/\Delta\nu\tau_{\it{int}}\right)^{1/2}$,
which is an estimate of the noise power in the reference cross-correlation, and
the prime indicates that $C_{\it{pc2}}$ has been approximated. The extra term
$\Psi^2$ stops the zero-mean fluctuations in $P^{}_{r_1r_2}$ from going too
close to zero, while for large $\it{INR}_r$ equation (\ref{MODIFIED SPECTRUM
MK2}) reduces to (\ref{CORRECTION SPECTRUM MK2}).

Since $\Psi^2$ introduces a small bias in a similar way to the MK1 cancellers,
a relation equivalent to (\ref{mk1 residual power}), but with a much smaller
bias, can be derived:

\begin{equation}
R_{\it{pc2}}^{\prime} 
  \approx \frac{G_{m}^{2}\sigma_I^2}
               {1+\it{INR}_1\it{INR}_2\Delta\nu\tau_{\it{int}}},
\label{stablised pc2 resid}
\end{equation}

This will have both a noise and a RFI component, as in (\ref{mk1 residual
power}), however, since the canceller is now biased like a MK1a canceller the
added receiver noise term will not average away. $R_{\it{pc2}}^{\prime}$
reduces as $\it{INR}^2$ multiplied by the number of samples in the time
average.

There is a similar problem for the pre-correlation MK2 adaptive cancellers. In
the lag domain, the division in (\ref{mk2 pre corr weights}) becomes a
multiplication by the inverse of a matrix with columns containing offset copies
of the $r_1$-$r_2$ cross-correlation function \citep{Widrow1985}. Divisions by
zero in the frequency domain due to interference-free frequency channels in
$P^{}_{r_1r_2}$ are manifest in the reference lag matrix as singular values.
One method of dealing with this is to use singular value decomposition to
decompose the matrix into two orthonormal triangular matrices and one diagonal
matrix (see $\S$ 2.9 of \citealt{Press1986}). Singular (or near-singular)
values can be selected when the relevant diagonal matrix elements are less than
a chosen threshold, such as $\Psi$. The singular parts of the matrix contain no
information about the correlated signal and are removed from the decomposition
matrices. The inverse matrix can then be reconstructed from the remaining parts
of the three decomposition matrices, and it will not function in the RFI-free
parts of the spectrum.

%%%%%%%%%%%%%%%%%%%%%%%%%%%%%%%%%%%%%%%%%%%%%%%%%%%%%%%%%%%%%%%%%%%%%%%%%%%%%%%%
%%%%%%%%%%%%%%%%%%%%%%%%%%%%%%%%%%%%%%%%%%%%%%%%%%%%%%%%%%%%%%%%%%%%%%%%%%%%%%%%
\section{SUMMARY}
\label{SUMMARY}

Interference cancelling using a single reference antenna can give excellent
results when the reference signal interference-to-noise ratio is large, and
there is more gain towards the interfering signal for the reference antenna
than for the astronomy antennae. However, receiver noise in the reference
signal means that a fraction of the interference will always remain after
cancelling. A second reference signal can be used to remove the noise bias and
give infinite interference attenuation, but a larger amount of reference
receiver noise is added during cancelling. For pre-correlation systems, a dual
canceller setup can be used to average the (zero-mean) receiver noise away, a
process that comes automatically with post-correlation cancellers. A breakdown
of the main properties for the different mitigation techniques is given in
table \ref{SUMMARY TABLE}.

It is important to note that even though the single-reference cancellers leave
residual RFI, the residual may be extremely small and well below the RMS noise.
This occurs when the interference-to-noise ratios of the reference signals are
very large, and the use of two references (in pre-correlation systems) might
just add complexity to the system with little or no benefit. However, if
maximum sensitivity is required, one should be aware that they will eventually
reach a non-zero residual signal.

Using two reference signals to remove the reference receiver noise bias removes
the inherent stability of the algorithms in situations where some or all of the
frequency channels are interference-free. Although there are applications in
which the passband will always be entirely filled with RFI (such as
observations in the GPS L1 and L2 bands), many interfering signals will only
take up a part of the band. In these cases the algorithms need a mechanism to
turn themselves off in the vacant frequency channels.

%\clearpage

%\parbox[t]{wd}
%\begin{tabular}{|m{1cm}|m{1.2cm}|m{1.2cm}|m{1.2cm}|m{3.65cm}|m{3.65cm}|m{1.6cm}|}

\begin{table*}
\center
\begin{tabular}{|c|c|c|c|c|c|c|}
\hline
%  &&&&&&\\
  &
  \parbox[t]{1.0cm}{Fig. no.} &
  \parbox[t]{0.8cm}{Ref. Ants.} &
  Filters &
  Configuration &
  Behaviour &
  \parbox[t]{1cm}{Eqn. no.} \\
  &&&&&&\\
\hline
%  &&&&&&\\
  MK1a &
  \ref{mk1 adaptive cancellers} &
  1 &
  1 &
  \parbox[t]{4.2cm}{One reference antenna feeds both filter and weight
  generator.} &
  \parbox[t]{4.2cm}{Minimises residual power, but includes some RFI. Turns off
  gracefully as $INR_r\rightarrow0$.} &
  \ref{mk1 residual power}, \ref{mk1 residuals a}, \ref{mk1 residuals b} \\
  &&&&&&\\
\hline
%  &&&&&&\\
  MK2a &
  \ref{mk2 adaptive cancellers} &
  2 &
  1 &
  \parbox[t]{4.2cm}{Independent reference antennae for filter and weight
  generator.} &
  \parbox[t]{4.2cm}{Totally cancels RFI, but adds extra noise. Must be
  controlled as $INR_r\rightarrow0$.} &
  \ref{mk2 residual power} \\
  &&&&&&\\
\hline
%  &&&&&&\\
  MK1b &
  \ref{indepRX MK1} &
  2 &
  2 &
  \parbox[t]{4.2cm}{Two MK1a cancellers, operating on copies of main signal.} &
  \parbox[t]{4.2cm}{Sensitivity decrease, and non-zero RFI floor.} &
  \ref{mk1 indepRX} \\
  &&&&&&\\
\hline
%  &&&&&&\\
  MK2b &
  \ref{indepRX MK2} &
  2 &
  2 &
  \parbox[t]{4.2cm}{Two MK2a cancellers, operating on copies of main signal.} &
  \parbox[t]{4.2cm}{Sensitivity decrease, but no non-zero RFI floor.} &
  \ref{mk2 indepRX} \\
  &&&&&&\\
\hline
%  &&&&&&\\
  PC2 &
  \parbox[t]{1cm}{not shown} &
  2 &
  1 &
  \parbox[t]{4.2cm}{Two reference correlations with each astronomy signal.} &
  \parbox[t]{4.2cm}{Sensitivity decrease, but no non-zero RFI floor. Must be
  controlled as $INR_r\rightarrow0$.} &
  \ref{pc2 residual power}, \ref{stablised pc2 resid} \\
  &&&&&&\\
\hline
\end{tabular}
\caption{Summary of the main properties for various interference mitigation
	 techniques\label{SUMMARY TABLE}}
\end{table*}

%\clearpage

%%%%%%%%%%%%%%%%%%%%%%%%%%%%%%%%%%%%%%%%%%%%%%%%%%%%%%%%%%%%%%%%%%%%%%%%%%%%%%%%
\acknowledgments

We are grateful to Dr. Mike J. Kesteven and Professor Lawrence E. Cram for
discussions and comments on this paper. The Australia Telescope Compact Array
is part of the Australia Telescope which is funded by the Commonwealth of
Australia for operation as a National Facility managed by CSIRO.


%% copy editing.  Individual instruments can be provided in parentheses,
%% after the keyword, but they will not be verified.

Facilities: \facility{ATCA}.

%%%%%%%%%%%%%%%%%%%%%%%%%%%%%%%%%%%%%%%%%%%%%%%%%%%%%%%%%%%%%%%%%%%%%%%%%%%%%%%%

%%%%%%%%%%%%%%%%%%%%%%%%%%%%%%%%%%%%%%%%%%%%%%%%%%%%%%%%%%%%%%%%%%%%%%%%%%%%%%%%

\end{document}